\documentclass[12pt,fullpage]{article}
\usepackage {epsfig}
\usepackage {amssymb}
\usepackage{rotating}

\def\mdot{{\raisebox{1pt}{\hbox{$\stackrel{\bullet}{M}$}}}\ \!\!}
\def\apgt{\ {\raise-.5ex\hbox{$\buildrel>\over{\scriptstyle\sim}$}}\ }
\def\aplt{\ {\raise-.5ex\hbox{$\buildrel<\over{\scriptstyle\sim}$}}\ }

\begin{document}

\title{\hbox to 16cm{\small\it УДК 524.387 + 524.338.3  \qquad\quad  PACS 97.80.G, 97.30.Q, 97.10.M \hfil}
\bigskip {\bf A Study of an Outburst in the Classical Symbiotic Star Z And
in a Colliding-Wind Model} }

\author{\bf D.V.~Bisikalo$^1$, A.A.~Boyarchuk$^1$, E.Yu.~Kilpio$^1$,\\
 \bf N.A.~Tomov$^2$, M.T.~Tomova$^2$\\[5mm]
$^1$ Institute of Astronomy, Moscow, Russia\\
$^2$ Institute of Astronomy, Rozhen National \\
Astronomical Observatory , Smolyan, Bulgaria\\}

\date{}

\maketitle

\begin{abstract}

Two-dimensional gas-dynamical modeling of the mass-flow structure
is used to study the outburst development in the
classical symbiotic star Z~And. The stage-by-stage rise of the
light during the outburst can be explained in the framework
of the colliding winds model. We suggest a scenario for the development of the
outburst and study the possible influence of the changes of the flow
structure on the light of the system. The model variations of
the luminosity due to the formation of a system of shocks are in
good agreement with the observed light variations.

\end{abstract}

\section{INTRODUCTION}

Mass transfer in symbiotic stars is realized because of the
existence of a  stellar wind of their cool giant component. Based
on radio data Seaquist and Taylor [\ref{Seaq90}] suggest that the
radio emission is determined by the wind of the giant, partially
ionized by the hot component. The mass-loss rate for Z And
obtained from radio data at 1.465 and 4.885 GHz in July 1982 by
Seaquist and Taylor [\ref{Seaq84}] amounts to  $\sim2\times10^{-7}
M_\odot$/yr and the velocity of the wind was estimated to be 25
km/s [\ref{FC88}].

The existence of the wind of the hot component of the symbiotic binaries during their active stages is
confirmed by the observations at least of some of them, for
example the system AG Peg [\ref{NV90}]. In 1989
Nussbaumer and Vogel [\ref{NV89}] suggested that the white dwarf
in Z And can also have a wind. The possibility for the existence of a wind
of the hot component of Z And during outburst phase was studied by
Nussbaumer and Walder [\ref{Nuss&W93}].

The observational data obtained during the 2000 -- 2002 outburst of Z And indicate the
presence of the white dwarf's wind in this system. It is revealed for instance by
observations of UV [\ref{Sokol02}, \ref{Sokol06}] and optical
[\ref{TTZ03}] lines. In November and December 2000 the
PV~$\lambda$1117 \AA\ UV line had a variable P Cygni profile with
one or two absorption components indicating outflow at velocities
of 0 -- 300 km/s. In all the cases, however, the maximal depth of the
absorption was at velocities of 0 -- 50 km/s [\ref{Sokol02}].
Both the optical and the UV data show an increase of the emission
component relative to the absorption one during the
outburst. The presence of a high velocity flow from the hot component during the light maximum was noted also by Skopal et al. [\ref{Skopal05}]. The analysis of the radio emission confirms the existence of
the white dwarf's wind during the outburst of Z And in 2000
[\ref{Sokol06}].

All of these data suggest that both the cool and the hot components have winds
in the active phases of classical symbiotic stars and their
behavior must be considered in the framework of the
colliding winds model.

The first attempts to describe qualitatively the behavior of
symbiotic stars supposing existence of a hot component's wind
during the outburst  were made in 1984  [\ref{Wills1}--%
\ref{Kwok}]. These studies were aimed to explain the observed line
profiles. Two-dimensional gas-dynamical modeling of colliding winds
was first undertaken in 1993--1996 by Nussbaumer and Walder
[\ref{Nuss&W93}] and Bisikalo et al. [\ref{Dima94}, \ref{Dima96}].

In the present paper we study the structure of the mass flow in
the framework of the two-dimensional gas-dynamical model when the
winds of the hot and the cool components present. A detailed
description of this model can be found in [18], where only the
donor's wind is considered. This model was modified - in
particular, in the computations where the hot component's wind was
taken into account, we used a spatially nonuniform $679\times589$
grid that becomes denser in the vicinity of the accretor. The use
of a denser grid in the vicinity of the accretor and parallel
computers enabled us to consider the features of the flow near the
accretor in detail. The wind of the accretor was modeled
introducing a jump of the parameters at its surface. The boundary
conditions were chosen in accordance with the observational data.
We considered different cases varying the  parameters of the model
(pressure, density, and velocity of the outflowing matter). The
boundary conditions (more precisely, their variations) were
selected to be consistent with the energy of the outburst. A
steady-state solution prior to the outburst was chosen as an
initial condition, the next computations were realized using
modified boundary conditions at the accretor's surface. We based
our modeling on the assumption that the wind of the hot component
appears after the onset of the outburst. We considered the
evolution of the conditions over a time span typical for the
outburst development, of the order of hundreds of days. The time
required to achieve a steady state  was not considered in detail
in the framework of our problem.

The results of the computations were used to explain the stage-by-stage
nature of the rise of the light during the development of the outburst. A
possible scenario for this development is
suggested.

\section{THE PECULIAR CHARACTER OF THE RISE OF THE Z And LIGHT DURING THE OUTBURST}

Z Andromedae is one of the most intensively observed symbiotic
stars. Several periods of activity have been detected over more
than a hundred years of observations -- after 1915, 1939, 1960, 1984,
and 2000. On average, the light of the system increases by
$2^m-3^m$ over a period of about 100 days during an outburst, after which it
begins to decline and returns to its initial value after
several hundred of days.

Detailed light curves were obtained during the last outburst,
which clearly show that the rise to the maximal light occurs in a
stage-by-stage mode. The light curves obtained during the previous
outbursts were not detailed enough to detect this effect.
Moreover, only for the two last active phases observational data
in different wavelength regions are available (obtained by
instruments on spacecraft).
\marginpar{\it\small\fbox{Fig.1}}

The UBV light curves for the 2000
outburst are shown in Fig.1. The first stage of the rise of the light started in the
end of August 2000 and continued for about 60 days. During that
time the light increased by $1.9^m$.
Further it remained constant one and even slightly
decreased over about 25--30 days. The light began to rise again
after November 13, 2000, and reached a second maximum (in fact,
the overall maximum of the outburst) after approximately 25 days
(near December 6, 2000). The times of the first and second
maxima are marked in Fig. 1 with dashed lines.

\marginpar{\it\small\fbox{Fig.2}}

The most detailed light curve,
published by Sokoloski et al. [\ref{Sokol06}] is shown in Fig.2. According to
these data, there is one feature more on the curve -- a kink that appears
about 2.5 weeks after the onset of the outburst. This led
Sokoloski et al. [\ref{Sokol06}] to distinguish three stages of
the rise of the light, separated by two plateaus. The rise stages
last 2.5, 2.5 -- 3, and slightly more than 3 weeks respectively,
while the first and the second plateaus last one week
and about one month.

The existence of the first maximum is explained in the works [\ref{Sokol06},
\ref{Sokol05}] as being due to clearing of the ejected shell and the future rise
of the light is related to the revealing the white dwarf. However, the
UV and the optical spectral data presented in [\ref{Sokol02}] and [\ref{TTZ03}]
contain P Cygni lines indicating mass
outflow from the dwarf at the time of the light maximum.
Thus, these  observational data cast some doubt on the suggestion
made in [\ref{Sokol06},\ref{Sokol05}].

In the framework of our investigation we have made some attempt to
explain the  behavior of the light of Z And using the
colliding-winds model.

 \section{DEVELOPMENT OF THE OUTBURST AFTER A WIND OF THE HOT COMPONENT APPEARS}

Thermonuclear burning at the surface of the accretor is considered
to be the most probable origin for the observed features of
symbiotic stars [\ref{Kenyon86}, \ref{IbenTu96}]. The burning of
hydrogen at the white-dwarf surface depends substantially on the
accretion rate [\ref{PR80}]. For a narrow range of accretion
rates stable hydrogen burning is possible
[\ref{Zytkow}--\ref{Fuji}]. For the mass of the white dwarf in Z And
$M=0.6M_\odot$ this range is $2.1\cdot10^{-8}M_\odot/\mbox{год}
\aplt\mdot^{accr}\aplt4.7\cdot10^{-8}M_\odot/\mbox{год}$.

In the commonly accepted picture the outbursts of classical
symbiotic stars realize when the accretion rate exceeds the upper
limit for stable hydrogen burning. If this occurs, the accreted
matter accumulates in a hydrogen-burning shell and expands to
giant dimensions. However, the scale of the observed outbursts
casts doubt on the possibility that they are associated purely
with accretion processes. The estimates show that the energetics of
the 2000 event require an accretion rate exceeding
$10^{-5}M_\odot/\mbox{yr}$ [\ref{Sokol06}], in contradiction with both
observations and computational results [\ref{Mitsumoto2005}].
We considered a ''combined'' mechanism of the outburst where
the increase of the accretion rate due to disruption of the disk
results in variations of the burning rate [\ref{Bisikalo2002},
\ref{Tutu76}--\ref{Kilpio_cefalu}]. In this case the amount of
the accreted matter ($10^{-8}- 10^{-7} M_\odot/\mbox{yr}$
[\ref{Mitsumoto2005}]) is sufficient to explain the observed
luminosity variations (according to the estimates given in
[\ref{Sokol06}], it is necessary to accrete $\sim
10^{-7}M_\odot$). A similar model was suggested by Sokoloski et
al. [\ref{Sokol06}], who supposed that the disk instability leads to
an increase of the accretion rate, which, from its side, causes
an increase of the rate of the nuclear burning.

The transition from quiescence to an active phase requires a
sufficient increase in the accretion rate. According to the
observational data, the mass-loss rate of the donor in Z And is
$\sim 2\times10^{-7}M_\odot/\mbox{yr}$. Since the amount of matter
accumulated in the disk during the inter outburst period
1997--2000 should not exceed $\sim 5\times10^{-7}M_\odot$, the
development of the outburst requires the accretion of the entire
disk even in the framework of the combined model where the increase of
the nuclear-burning rate is taking into account. As a rule, accretion-disk
instabilities result in the infall of $\sim 10\%$ of the total
mass of the disk, which is clearly not enough.

The required increase of the accretion rate can be provided in the framework of
the mechanism suggested by us in [\ref{Bisikalo2002}], based on
the results of two-dimensional gas-dynamical modeling and
confirmed by three dimensional computations in our work
[\ref{Mitsumoto2005}] as well. According to that mechanism even a small increase of the velocity of the donor's wind is sufficient to change the accretion regime. At the time of the transition from disk acretion regime to wind one, the disk disrupts and the wind with increased velocity causes the infalling its material onto the surface of the accretor. The alalysis of the results of these computations has shown the following points.

\begin{enumerate}

  \item Variation of 5 km/s of the observed donor's wind velocity of 25 km/s
  [\ref{FC88}] is sufficient to change significantly the flow pattern as well as the accretion regime - from disk accretion to wind one. When the wind velocity amounts to 30 km/s, a conical shock forms. When it is equal to 20 km/s, an accretion disk appears in the system.

  \item In the quiescence the accretion rate is $\sim (4.5-5)\times10^{-8}M_\odot/\mbox{yr}$
(for a mass-loss rate of the donor of $\sim
2\times10^{-7}M_\odot/\mbox{yr}$  [\ref{FC88}]) which corresponds
to the range of stable hydrogen burning for the adopted mass of
the white dwarf in Z And.

  \item If the wind velocity increases from 20 km/s to 30 km/s, the disk will
  disrupt and its  material will fall onto the surface of the accretor.

 \item The disruption of the disk is accompanied by a jump in the
accretion rate (Fig. 3).
\marginpar{\it\small\fbox{Fig.3}}
A growth of the accretion rate by a factor of $\sim 2.0-2.2$
relative to its initial value provides exceeding the upper
limit of the region of stable hydrogen burning. According to our computations
the disk is fully disrupted in about 100 days and a mass of several units of
$10^{-8}-10^{-7}M_\odot$ accretes during that time.

\end{enumerate}

This amount of matter turns out to be sufficient to increase the pressure and
the temperature providing an increase of the nuclear-burning
rate. According to [\ref{Sokol06}] the typical time scale for the
response of the nuclear-burning shell after the accretion of a
sufficient amount of matter is about one month, which is in good agreement
with the observed time of the formation of a kink of the light
curve. This time is 2.5 weeks after the onset of the outburst.
The growth of the luminosity during the time interval
before the increase of the burning rate
is due to rise of the accretion rate. It means  the time
scale of that rise must be in agreement with that of the increase
of the luminosity to the first kink of the light
curve. In our computations the disk was fully disrupted in about
100 days. The inconsistency between the time scales can be
compensated taking into account the  nonuniform variation of the
accretion rate. Figure 3 shows the accretion rate rises
very rapidly and reaches its maximum in 10--20 days. If we suppose that the amount
of matter accreted during that time is sufficient to initiate an
enhanced burning rate, the further increase of the luminosity will
be due to both the continuing accretion and the enhanced burning
rate. According to [\ref{Sokol06}] after the appearance of the first kink of the
light curve associated with the enhancement of the burning rate,
an expanding envelope -- a pseudophotosphere and/or optically
thick wind forms in the system.

It is commonly accepted that the increase of the visual light is due to the energy redistribution towards
the longer wavelengths during the expansion of the envelope of the compact
component [\ref{Sokol02}, \ref{Tomov03}, \ref{FC95}]. However,
an expansion of the envelope on its own is unable to explain the
above features of the light curve. Moreover, the presence of the
wind in the system should influence its brightness. If the wind
of the white dwarf arises as the outburst progresses (after 2.5 weeks
after the beginning of the outburst according to the data in[\ref{Sokol06}]),
it will seem that its influence
will begin to manifest itself not at the very
beginning of the outburst, but in its later stages.

Since the hot wind arises during the outburst, the curve
of the optical light must be a sum
of three components:

\begin{enumerate}

  \item [1)] a luminosity variation causing an  expansion of the pseudophotosphere
  (the increase of the flux of the nebula due to additional radiative ionization
  can be neglected, since it is proportional to the Lyman luminosity and does not
  change the shape of the curve);

  \item [2)] a light increase due to the wind propagation in the nebula;
  at that the inflowing matter has the temperature of the hot wind, resulting in the
  formation of a high-temperature region in the nebula;
  \item [3)] a luminosity increase due to the formation of shock structures
  appeared as a result of wind collision.

\end {enumerate}

If the effects related to the wind are strong enough, they should be
visible in the light curve. It is evident that the increase of the luminosity
of the hot component and the expansion of its
pseudophotosphere have the main contribution in the variation of the visual
light. If at some time the component determined by the
wind propagation in the nebula is added, followed by the formation
of shocks providing further brightness increase, the resulting
light curve will have two variations of its slope related to the
hot wind.

The shape of the light curve will be more complex if the flux of
the expanding photosphere has time variations. It
is obviously that, at some time, the expansion of the
photosphere will be replaced by a contraction. This time
will correspond to a peak in the light curve. Its final shape
depends on the time of the hot wind appearance (the second term),
the time when the shocks form (the third term) and the position of the peak.

The increased accretion rate due to the disruption of the disk
typically lasts for about 100 days. It means that the expansion
of the envelope will stop after about 100 days, i.e. its flux (the first term)
will begin to decrease, giving rise to a peak in the time interval
$0^{\rm d}-100^{\rm d}$ in the light curve. The wind appears in this time
interval too (after $\sim 20^{\rm d}$ after the onset of the outburst
[\ref{Sokol06}]). Starting with this time, the contribution of the
high temperature region of the nebula to the luminosity of the system will be
positive till the epoch of the wind disappearance. The velocity of the wind is about 50
km/s. Then, assuming the shock is located between the
components of the system (at a distance of $1/3\, A - 2/3 \,A$
from the accretor, where $  A  $ is the component's separation), we
can estimate the time that elapses before the contribution of
the shock to become substantial. This time turns out to be 26-- 52
days after the onset of the wind. The peak determined by the contribution
of the shock will be attained after some time in addition.

To estimate the time scale of the shock's development we carried
out a set of computations with parameters close to the conditions
for the Z And outburst. The results show that after the change of
the conditions at the accretor's surface an S-shaped shock structure
and contact discontinuity form in the space between the
components.

Figure 4
\marginpar{\it\small\fbox{Fig.4}}
shows the development of events after the change of conditions at
the accretor's surface. In the model presented the wind velocity
at the accretor's surface is 50 km/s. The density distribution and
the velocity vectors for the entire computational domain are shown
for six times corresponding to 12, 42, 70, 103, 125, and 150 days
after the start of the computations. Shocks are seen as
concentrations of the density contours. The orbital motion of the
accretor is counterclockwise. The dashed lines show the contours
of the standard Roche potential. It is seen that the system of
shocks forms fairly quickly -- the shock occupies its final location
on the X axis between the components already after about 70 days
(Fig. 4c). The analysis of the results shows that the shock is
established first between the components and in the regions above
and below the line of their centers it forms definitively 20 -- 30
days later (Fig. 4d). Further the parameters of the shock do not
change substantially (Figs. 4d--4f).

Summarizing the results of the numerical modeling it must be expected that
the first component (the expansion of the
pseudophotosphere) of the light curve will reach its maximum in the time interval
from $0^{\rm d}$ to $100^{\rm d}$. The contribution of the second
component will begin to be significant in the same time interval
and 70 days after that the third component will begin to make its contribution too.
It will reach its maximum after 20 -- 30 days in adition.

In our analysis of the light curve of the 2000 outburst of Z And
we can assume that the first maximum, when the light rose by $1.9^m$ in $\sim 60$ days, is determined only by
the expansion of the
optically thick pseudophotosphere (the first term). It is known
from observations [\ref{Sokol06}] that the first kink of the light
curve appeared close to September 15, 2000, and is
related to the onset of the wind of the hot component.
Assuming that the time of the wind appearance is close to
September 15, 2000, it can be expected , according to our
computations, that for a wind velocity of 50
km/s the shock will begin to form 70 days after
the rise of the first kink, i.e. after the appearance of
the wind. The analysis of the light curve shows that the second
change of its slope realizes on November 13, 2000, i. e. after 60 days.
The peak associated with the formation of the shock should be
shifted at 20 -- 30 days (according to the computations), while the
observational data show that it forms after 25 days (December 6,
2000). Thus, the comparison of the results of the computations with
the observations shows all main stages of the change of the
light are in good agreement with the model.

To be sure that the presence of the wind (the second term)
and the shock (the third term) really provide the observed
light variations we must estimate their contributions.

The contribution of the hot wind to the light of the system can
be estimated assuming that the location of the shock -- the boundary of
the areas of the two winds, is determined
by the condition for equality of the ram pressures: $\rho_1 v_1^2 = \rho_2 v_2^2$.
We can estimate the density of the hot
wind assuming that the boundary is located in the middle of
the components' separation (Fig. 4). Using the known
parameters of the donor's wind (a mass-loss rate of $\sim
2\times10^{-7}M_\odot/\mbox{yr}$, a velocity of 30 km/s) and
the velocity of the hot wind (50 km/s), we can estimate
the mass-loss rate of the hot component. It turned out to be $\sim
1.2\times10^{-7}M_\odot/\mbox{yr}$.

Assuming that the hot wind is spherically symmetric we can
calculate its $UBV$ fluxes emitted by a spherical layer.
The inner boundary of this layer is equal to $2.2 R_\odot$ and is close to the observed radius of the pseudophotosphere on November 22, 2000 and the outer one of $240.5 R_\odot$ is equal to the half of the component separation. The emission
of the region beyond the outer edge can be neglected. It was supposed in our calculations that the gas consists of hydrogen and ionized
helium (the wind region is hot), the helium abundance is 0.1, and
the distance to the system is 1.12 kpc. The
contribution of the wind in the $UBV$ bands is shown in Tabl.1. It is seen that it is fairly large being on average $\sim20\%$ on November 22, 2000.

\marginpar{\it\small\fbox{Table 1}}

In regard to the contribution of
the shock, according to our computations we can present the following view.
A calculated structure of the wind
collision is presented  in Fig. 5. The notation is similar to that in Fig. 4.
The region in the vicinity of the shocks is shown in Fig.6 in more detail.

\marginpar{\it\small\fbox{Fig.5}}

\marginpar{\it\small\fbox{Fig.6}}

Our analysis shows that the region between the shocks has a
considerably higher temperature than the surrounding medium: on
the X axis passing through the component centers, it is
higher by a factor of 50 than in the surrounding
regions of the nebula and reaches $10^6$ K. This is in agreement
with the results of Nussbaumer and Walder [\ref{Nuss&W93}] for
their model of a symbiotic star with colliding winds. It was
suggested in their work that just the high-temperature region
between these shocks is the source of the X-ray emission observed
from some symbiotic stars. Our estimates based on the method
described in [\ref{G&W}] show that for the assumed parameters of
the winds and the computed area of the shock the X-ray luminosity is
$10^{31} - 10^{32}$ erg. This result is in very good agreement with the observed
X-ray luminosity during the 2000 outburst of Z And
[\ref{Sokol06}].

To estimate the contribution of the shocks in the total luminosity
of the system we assumed that the light variations are
proportional to the energy losses in the system. We made an
approximation adopting that the radiative losses per unit time per unit
volume are proportional to $\rho^2\Lambda(T)$, where
$\Lambda(T)$ is the cooling function [\ref{Cox}]. The total
energy loss in the system was estimated as a sum over all cells in the computational domain. By means of computing the total energy
loss for the time when the system of shocks was formed and
comparing it with the energy loss in quiescence we can estimate
the change of the light of the system. Our analysis shows
that the quantity $\sum\rho^2\Lambda(T)$ can be appreciably higher after the
formation of the system of shocks than in quiescence and the
system of shocks can increase the light of the system by
$1^m$.

Summarizing our results we conclude that both the wind (the
second term) and the shock (the third term) can significantly
change the light of the system. Since the appearance of
these phenomena in the framework of the model is at the same time as  observed, we
suggest that precisely they are responsible for the observed
behavior of the light.

\section{COMPARISON WITH OBSERVATIONS AND ANALYSIS OF
THE CONTRIBUTION OF THE SHOCK TO THE TOTAL
LIGHT OF THE SYSTEM}

In the previous section we showed that the proposed scenario of
the outburst based on our modeling is in good agreement with
the observed temporal behavior of the light. If this model
is correct, the presence of the shock will also result in other
observable effects, for example, such as existence of shock ionization.
Let us consider some rezults of the
observations of the Z And outburst.

According to the observational data [\ref{Sokol06}, \ref{Tomov03}],
the last stage of the rise of the light started on November 13,
2000 (JD 2451862) and lasted for about 25 days. In the proposed
scenario this stage is provided by formation of a system of
shocks resulting from the collision of the winds. To
estimate the influence of these shocks on the light of the system
we used the results from the work of Tomov et al. [\ref{Tomov03}], where
the basic parameters of the system's components
(the cool component, the hot component, and the nebula) and their continuum fluxes were estimated from observational data.

The continuum
\marginpar{\it\small\fbox{Table 2}}
fluxes of the components of Z And in units of
$10^{-12}$ erg$\cdot$cm$^{-2}$s$^{-1}$\AA$^{-1}$ for November 22
and December 6, 2000 [\ref{Tomov03}] are listed in Table 2.
These data show that the emission of both the
nebula and the hot component increased during this time interval.
We used the difference between the total (model) and the observed
fluxes as a percentage of the observed flux as a criterion for the
agreement between the model continuum and the observed one. On the other hand,
the UV continuum fluxes at wavelengths
1059 and 1103 \AA\ where only the hot component emits, taken from the work of Sokoloski et al.
[\ref{Sokol06}], show that its emission increased during the period
November 16 -- November 27, 2000, and was constant after that, till December 15, 2000. The result of the continuum analysis is in
qualitative agreement with the UV data, as far as the time interval
November 22 -- December 6 covers the interval November 22
-- 27 when the emission of the hot component was rising.
However, since this emission was not changing between November 27
and December 15, we propose a second variant of the continuum  analysis for December 6 too in Table 2, with a smaller growth of
the radius of the hot component. When we calculate the emission due to shock ionization we shall consider this second variant too,
since it is in better agreement with the observed behavior of the
hot component in the UV, although the total
and the observed fluxes in the infrared are in better agreement in the first variant.

According to the proposed model three processes contribute to the light
curve: the rise of luminosity causing an expansion of the envelope, the rise and
development of the high temperature region in the nebula formed
by the white dwarf's hot wind, and the shock structure formed by
the collision of the winds.

The observed $UBVRIJHKLM$ fluxes at different times are shown in
Fig.7. These times are as follows: quiescence, the onset of the
last stage of the growth of the light (November 13, 2000), some
time of the rise to the second maximum (November 22, 2000) and
some time close to the light maximum (December 6, 2000).
\marginpar{\it\small\fbox{Fig.7}}
These data show that the flux at shorter wavelengths grows more when the optical
light goes toward its second maximum. Most probably it is due to the
appearance of the hot wind and the formation of the shocks.

The most correct approach to determine the contribution of the shock is to
estimate the shock ionization. Let us consider the ratio of the number of
ionizing photons and the number of recombinations in the
nebula $\mu$ in the quiescence and on November 22 and December 6, 2000. This
ratio is estimated as
\[
\mu = {\frac{{L}}{{n_{e} n_{ +}  \alpha V}}}\,\,{\rm ,}
\]
where $L$ is the Lyman photon luminosity of the hot component,
$n_{e}$ and $n_{ +} $ are the number densities of the electrons
and ions respectively, $\alpha $ is the total (to all levels)
recombination coefficient, and $V$ is the volume of the nebula.
The Lyman luminosity is
\[
L = 4\pi R^{2}H_{\lambda < 912} = 8\pi
^{2}{\frac{{R^{2}}}{{c^{2}}}}\left( {{\frac{{kT}}{{h}}}}
\right)^{3}G(T)\,\,{\rm ,}
\]
where $R$ and $T$ are the radius and effective temperature, $G(T)$
is a function related to the number of the ionizing photons (given in
numerical form in the book of Pottasch [\ref{Pottasch87}]), and
the remaining quantities have their commonly accepted meaning. The
number of recombinations is
\[
n_{e} n_{ +}  \alpha V = \left[ {1 + a({\rm He})} \right]\alpha
n^{2}V\,\,{\rm ,}
\]
where \textit{a{\rm (He)}} is the number abundance of helium
relative to hydrogen. If we obtain from observations the radius and the
effective temperature of the hot component and the emission measure of the nebula
in addition, we will calculate the ratio $\mu $.

In the state of ionization equilibrium when only radiative
ionization is realized in the nebula $\mu  \ge $1. The equality
is satisfied when all photons are absorbed in the nebula. When
$\mu $<1 the number of recombinations is greater, which means
that shock ionization is realized in the nebula
along with radiative one. The ratio of the continuum fluxes
due to shock and radiative ionization is $(1-\mu  )/ \mu
$  when all photons are absorbed in the nebula. In some cases of
the distribution of the circumstellar gas some
fraction of the photons can leave the nebula. Then $(1-\mu  )/ \mu
$  is a lower limit of the ratio of the parts of the
continua corresponding to shock and radiative ionization.

The ratio $\mu $ was calculated with use of the parameters of the
system's components determined for different times during the
outburst in the work of Tomov et al. [\ref{Tomov03}]. These
authors found that the dominant ionization state of helium in the
nebula in quiescence is He$^{++}$ and during the outburst --
He$^{+}$. It was assumed that the nebular continuum in quiescence
is emitted by hydrogen and ionized helium and during the outburst
--
 by hydrogen and neutral helium.
These assumptions were taken into account when
computing $\mu $. For this purpose the value of \textit{a}({\rm
He}) was doubled for quiescence. The helium abundance
was taken to be 0.1, in accordance with the results of Nussbaumer
and Vogel [\ref{NV89}]. It was assumed that the average number
density in the Z And nebula is $10^{8}-10^{10}$ cm$^{-3}$
[\ref{FC88}, \ref{Proga94}--\ref{Birriel98}]. The recombination
coefficients were taken from [\ref{SH95}] for Menzel case B.

The computational results and their rms errors are presented in
Table 3.
\marginpar{\it\small\fbox{Table 3}}
The errors were derived from the uncertainties of the parameters
of the system given in [\ref{Tomov03}]. We did not present the
error for the quiescence, since in this case the Lyman luminosity
was calculated using the average temperature from the results of other authors,
based on UV data. We present both variants for December 6, 2000.

The data in Table 3 show that $\mu>1$ in quiescence
(September 15, 1999). It is known that the Z And nebula is
partially ionized in quiescence [\ref{FC88}, \ref{MK96},
\ref{Birriel98}, \ref{SS97}] and our result (within the errors)
indicates that some fraction of the photons leave the ionized
region. The ratio $\mu $ is equal to unity for November 22, 2000 (less than in
quiescence) and less than unity for December 6, 2000. Thus it decreases
with time, which means that the role of the shock ionization increases.

The second variant for the time of the light maximum proposes $\mu=0.76$. This means
there is no doubt that the shock ionization takes place and its
contribution to the emission of the nebula is not less than 0.24
(since some fraction of the photons can leave the nebula). This
contribution can be as high as 0.44 when we take into account the
observational uncertainties. The $UBV$ fluxes determined by the shock ionization
in the case of $(1-\mu) = 0.24$ are presented in Table 4.
The contribution of the wind is presented also in this table.
\marginpar{\it\small\fbox{Table 4}}
This contribution is the same for the two epochs since the parameters of the
wind are considered to be constant and, according to the model (Fig.
4), the boundary between the winds does not
change after some time. Subtracting the contribution of the hot
wind and the shock from the total emission of the nebula, we find that
the flux of its cool part changes because of the increase of the radiative
ionization during the period November 22 -- 27, 2000 resulting from
the growth of the hot component's luminosity. This
change, however, is insignificant.

Thus, our analysis of the observational data shows that the
light variations at the last stage of the outburst's
development are well explained by means of the proposed scenario and
consequently can be interpreted in the framework of the model of the colliding winds.

\bigskip

\section{CONCLUSION}

\bigskip

We have used the results of gas-dynamical modeling of flow
structures to study the development of the outburst in the
symbiotic system Z And.

The analysis of the Z And outburst in 2000
shows that the accretion processes are not able to provide the observed
energetics of that event. As a possible mechanism of the outburst's
development we considered a combined case when
the increase of the accretion rate as a result of
the disruption of the disk leads to variation of the burning rate.
This model was proposed in [\ref{Bisikalo2002},
\ref{Tutu76}--\ref{Kilpio_cefalu}] where it was assumed that the
variations of the velocity regime of the donor's wind result in disruption of
the accretion disk and the infall of a considerable amount of its
matter ($10^{-8}  - 10^{-7} M_\odot$ according to
[\ref{Mitsumoto2005}]) onto the surface of the white dwarf. This
is enough to initiate an increase of the rate of the nuclear
burning and, consequently, the subsequent increase of the luminosity (the
development of the outburst) will be determined by both the
ongoing accretion and the increased nuclear-burning rate.

According to Sokoloski et al. [\ref{Sokol06}] who analyze a
similar combined outburst model, an expanding envelope
(pseudophotosphere) and/or optically thick wind form in the system
after the first kink of the light curve, which is associated with
an enhancement of the nuclear burning. The presence of the wind in the Z
And system during its 2000 outburst is confirmed by numerous
observations in the UV, optical and radio regions. In this case the curve
of the optical light will be formed by (1)
a luminosity variation leading to an expansion of the
pseudophotosphere, (2) a light increase due to the wind
propagation in the nebula, (3) a luminosity increase
due to the formation of shock structures appeared as a result of wind collision.
As it is seen from the results of the computation of the gas-dynamical
structure, the effects associated with the wind are fairly
strong and just they determine the complex stage-by-stage nature of
the rise of the light during the period when the outburst progresses.

The proposed scenario for the development of the outburst provides
an explanation of the behavior of the light, which is consistent
with the available observational data -- it is in accordance with
the observed temporal characteristics, amplitudes and scale of the
shock ionization.

\medskip
This work was supported by the Russian Foundation for Basic
Research (project codes 05-02-16123, 05-02-17070, 5-02-17874,
06-02-16097, 06-02-16234), the Program of Support for Scientific
Schools of the Russian Federation (NSh-4820.2006.2), and the
Programs of the Presidium of the Russian Academy of Sciences
"`Origin and Evolution of Stars and Galaxies"' and "`Fundamental
Problems of Informatics and Informatics Technologies"'.

\newpage

\section*{REFERENCES}

\begin{enumerate}

\item
\label{Seaq90} Seaquist~E.~R., Taylor~A.~R., Astrophys J. {\bf
349}, 155 (1990).

\item
\label{Seaq84} Seaquist~E.~R., Taylor~A.~R., Button S., Astrophys
J. {\bf 284}, 202  (1984).

\item
\label{FC88} T.~Fernandez-Castro, A.~Cassatella, A.~Gimenez, {\it
et al.}, Astrophys. J. {\bf 324}, 1016 (1988).

\item
\label{NV90} Nussbaumer~H., Vogel~M.,  Astron. Gesellschaft
Abstract Ser. {\bf 4}, 19 (1990).

\item
\label{NV89} Nussbaumer~H., Vogel~M., Astron. and Astrophys.
{\bf 213}, 137 (1989).

\item
\label{Nuss&W93} H.~Nussbaumer, R.~Walder, Astron. and Astrophys.
{\bf 278}, 209 (1993).

\item
\label{Sokol02} J.~L.~Sokoloski, S.~J.~Kenyon, {\it et al.}, in
{\it The Physics of Cataclysmic Variables and Related Objects},
 eds  B.-T. Gansicke, K.
Beuermann and K. Reinsch,  ASP Conf. Proc. {\bf 261}  (San
Francisco: ASP, 2002), p. 667.

\item
\label{Sokol06} Sokoloski J.L., Kenyon S.J., Espey B.R., {\it et
al.}, Astrophys. J. {\bf 636},   1002 (2006).

\item
\label{TTZ03}   Tomov, N. A., Tomova, M. T., and Zamanov, R. K.
  in {\it Symbiotic Stars Probing Stellar Evolution}, eds
R. L. M. Corradi, J. Mikolajewska, and T. J. Mahoney,  ASP Conf.
Ser. {\bf 303} (San Francisco: ASP, 2003), p. 254.

\item
\label{Skopal05} Skopal~A., Errico~L., Vittone~A.~A., {\it et
al.}, in {\it Interacting Binaries: Accretion, Evolution, and
Outcomes}, eds  L.A. Antonelli, L. Burderi, F. D'Antona, T. Di
Salvo, G.L. Israel, L. Piersanti, O. Straniero, A. Tornambe,  AIP
Conf. Proc. {\bf 797} (2005),  p.~557.

\item
\label{Wills1} Willson~L.A., Wallerstein~G., Brugel~E.W., Stencel
~R.E., Astron. and Astrophys. {\bf 133},  154 (1984).

\item
\label{Wills2} Willson~L.A., Salzer~J., Wallerstein~G., and
Brugel~E., Astron. and Astrophys. {\bf 133},   137 (1984).

\item
\label{G&W} Girard~T., Willson~L.~A., Astron. and Astrophys. {\bf
183},   247 (1987).

\item
\label{Kwok&Leahy} Kwok~S., Leahy~D.~A., Astrophys.J. {\bf 283},
  675 (1984).

\item
\label{Kwok} Kwok~S., in {\it The Symbiotic Phenomenon}, eds
J.~Mikolajewska, M.~Friedjung et al., Proc. IAU Coll. No. 103,
Astrophys. and Space Sci. Library. {\bf 145} (Dordrecht, Kluwer
Academic Publishers, 1988), p.~129.

\item
\label{Dima94} D.~V.~Bisikalo, A.~A.~Boyarchuk, O.~A.~Kuznetsov,
{\it et al.}, Astron. Zh. {\bf 71}, 560 (1994) [Astron. Rep. {\bf
38}, 494 (1994)].

\item
\label{Dima96} D.~V.~Bisikalo, A.~A.~Boyarchuk, O.~A.~Kuznetsov,
and V. M. Chechetkin, Astron. Zh. {\bf 73}, 727 (1996) [Astron.
Rep. {\bf 40}, 662 (1996)].

\item \label{Bisikalo2002} D.~V.~Bisikalo, A.~A.~Boyarchuk, E.~Yu.~Kilpio, and
O.~A.~Kuznetsov, Astron. Zh. {\bf 79}, 1131 (2002) [Astron. Rep.
{\bf 46}, 1022 (2002)].

\item
\label{Sk02} Skopal~A., Vanko~M., Pribulla~T., {\it et al.}
Contrib. Astron. Observ. Skalnate Pleso {\bf 32}, 62 (2002).

\item
\label{Sk04}    Skopal~A., Pribulla~T., Vanko~M., {\it et al.}
Contrib. Astron. Observ. Skalnate Pleso {\bf 34}, 45 (2004).

\item
\label{Sokol05} J.~L.~Sokoloski S.~J.~Kenyon, {\it et al.}, in
{\it The Astrophysics of Cataclysmic Variables and Related
Objects},  eds J.-M. Hameury and J.-P. Lasota, ASP Conf. Ser. {\bf
330} (San Francisco: ASP, 2005), p. 293.

\item
\label{Kenyon86} S.~J.~Kenyon, {\it The Symbiotic Stars}
(Cambridge: Cambridge Univ. Press,  1986).

\item
\label{IbenTu96} I.~Iben,~Jr. and A.~V.~Tutukov, Astrophys. J.
Suppl. Ser. {\bf 105},   145 (1996).

\item
\label{PR80} B.~Paczy\'nski and B.~Rudak, Astron. and Astrophys.
{\bf 82},  349 (1980).

\item
\label{Zytkow} B.~Paczy\'nski and A.~\.Zytkow, Astrophys. J. {\bf
222},   604 (1978).

\item
\label{Sion} E.~M.~Sion, M.~J.~Acierno, and S.~Tomczyk, Astrophys.
J. {\bf 230},  832 (1979).

\item
\label{Fuji} M.~Y.~Fujimoto, Astrophys. J. {\bf 257},  767 (1982).

\item
\label{Mitsumoto2005} M.~Mitsumoto, B.~Jahanara, T.~Matsuda, et
al., Astron. Zh. {\bf 82}, 990 (2005) [Astron. Rep. {\bf 49}, 884
(2005)].

\item
\label{Tutu76} A.~V.~Tutukov and L.~R.~Yungel’son, Astrofizika
{\bf 12},  521 (1976).

\item
\label{Kilpio_strsbg} Kilpio~E.~Yu., Bisikalo~D.~V.,
Boyarchuk~A.~A., and Kuznetsov~O.~A., in {\it The Astrophysics of
Cataclysmic Variables and Related Objects}, eds J.-M. Hameury and
J.-P. Lasota, ASP Conf. Ser. {\bf 330} (San Francisco: ASP, 2005),
p. 457.

\item
\label{Kilpio_cefalu} Kilpio~E.~Yu., Bisikalo~D.~V.,
Boyarchuk~A.~A., and Kuznetsov~O.~A.,  in {\it Interacting
Binaries: Accretion, Evolution, and Outcomes}, eds  L.A.
Antonelli, L. Burderi, F. D'Antona, T. Di Salvo, G.L. Israel, L.
Piersanti, O. Straniero, A. Tornambe,  AIP Conf. Proc. {\bf 797}
(2005),  p.~573.

\item
\label{Tomov03} Tomov~N.A., Taranova~O.G., and  Tomova~M.T.,
Astron. and Astrophys. {\bf 401},   669 (2003).

\item
\label{FC95} T.~Fernandez-Castro, R.~Gonzales-Riestra, and
A.~Cassatella, {\it et al.}, Astrophys. J.  {\bf 442}, 366 (1995).

\item
\label{Cox} Cox~D.~P.  and Daltabuit~E., Astrophys. J. {\bf 167},
  113 (1971).

\item
\label{Pottasch87} S.~R.~Pottasch, {\it Planetary Nebulae – A
Study of Late Stages of Stellar Evolution}, (Reidel, Dordrecht,
1984), 335 p.

\item
\label{Proga94} D.~Proga, J.~Mikolajewska, and S.~J.~Kenyon,
Monthly Not. Roy. Astron. Soc. {\bf 268},  213 (1994).

\item
\label{MK96}    Mikolajewska~J. and Kenyon~S.~J., Astron. J. {\bf
112},  1659 (1996).

\item
\label{Birriel98}   Birriel~ J.~J., Espey~B.~R.,
Schulte-Ladbeck~R.~E., Astrophys. J. {\bf 507}, L75 (1998).

\item
\label{SH95} P.~J.~Storey and D.~G.~Hummer, Monthly Not. Roy.
Astron. Soc.  {\bf 272},   41 (1995).

\item
\label{SS97} Schmid~H.~M. and Schild~H., Astron. and Astrophys.
{\bf 327},   219 (1997).

\end{enumerate}

\clearpage
\begin{center}

\begin{table}
\caption {Emission of the hot wind}
\bigskip
\begin{tabular}{|c|c|c|}
\hline
  Photometric & Flux, & Fraction of the nebula emission\\
  waveband &  $10^{-12}$~erg$\cdot$cm$^{-2}$s$^{-1}$\AA$^{-1}$&  on November 22.11.2000 \\
\hline
  $U$ & 0.4646 & 23$\%$ \\
  $B$ & 0.1346 & 19$\%$ \\
  $V$ & 0.1180 & 19$\%$ \\ \hline
\end{tabular}
\end{table}

\begin{sidewaystable}
\caption{Continuum fluxes of the Z And components (in
10$^{-12}$~erg$\cdot$cm$^{-2}$s$^{-1}$\AA$^{-1}$)
[\ref{Tomov03}]}
\bigskip
\begin{tabular}
{|p{4.5cm}|p{1cm}|p{1cm}|p{1cm}|p{1cm}|p{1cm}|p{1cm}|p{1cm}|p{1cm}|p{1cm}|p{1cm}|}
\hline
Emission source .$^*$ $^{{\rm а}}$\textsf{}& $U$& $B$& $V$&
$R$& $I$& $J$& $H$& $K$& $L$&
$M$ \\
\hline
\multicolumn{11}{|c|}{November 22 2000 \quad ($R_{{\rm h}{\rm o}{\rm t}} = 2.22 R_{\odot}$, $n^{{\rm 2}} V = 17.4\times 10 ^{{\rm 5}{\rm 9}}$ cm$^{-3}$)}\\
\hline
Cool component& 0.020& 0.160& 0.376& 0.710& 1.755& 1.343&
0.856& 0.439& 0.113&
0.034 \\
\hline
Hot component& 5.537& 2.934\textsf{}& 1.336& 0.730& 0.340& 0.063&
0.022& 0.007& ~~~--&
~~~-- \\
\hline
Nebula& 1.983& 0.717& 0.631\textsf{}& 0.547& 0.432& 0.147& 0.081&
0.051& 0.020&
~~~--\textsf{} \\
\hline
  Total flux  &
7.540&
3.811&
2.343&
1.987&
2.527&
1.553&
0.959&
0.497&
0.133&
0.034 \\
\hline
Observed flux  & 7.205& 3.848& 2.492& ~~~--& ~~~--& 1.591& 0.927&
0.461& 0.117&
0.024 \\
\hline
Observational error& 0.054& 0.036& 0.022& ~~~--& ~~~--& 0.002&
0.001& 0.001& 0.001&
0.001 \\
\hline
Deviation $D $ & 5\%& -1\%& -6\%& ~~~--\textsf{}& ~~~--\textsf{}&
-2\%& 3\%& 8\%& 14\%&
42\% \\
\hline
\multicolumn{11}{|c|}{December 6 2000  -- variant 1 \quad ($R_{{\rm h}{\rm o}{\rm t}} = 2.36 R_{\odot}$ , $n^{{\rm 2}} V = 20.9\times 10 ^{{\rm 5}{\rm 9}}$ cm$^{-3}$)}\\
\hline Cool component& 0.020& 0.160& 0.376& 0.710& 1.755& 1.343&
0.856& 0.439& 0.113&
0.034 \\
\hline
Hot component & 6.257& 3.315& 1.510\textsf{}& 0.826&
0.384\textsf{}& 0.071\textsf{}& 0.024& 0.008& ~~~--&
~~~-- \\
\hline
Nebula& 2.382& 0.861& 0.758& 0.657& 0.519& 0.176& 0.098& 0.061&
0.024&
~~~--\textsf{} \\
\hline
Total flux& 8.659\textsf{}& 4.336\textsf{}& 2.644\textsf{}& 2.192&
2.658& 1.590\textsf{}& 0.978\textsf{}& 0.508\textsf{}&
0.137\textsf{}&
0.034 \\
\hline
Observed flux& 8.662& 4.257& 2.682& ~~~--& ~~~--& 1.635& 0.944&
0.478& 0.120&
0.024 \\
\hline
Observational error& 0.064& 0.039& 0.023& ~~~--& ~~~--& 0.001&
0.001& 0.001& 0.001&
0.001 \\
\hline
Deviation $D  $ & 0\%\textsf{}& 2\%\textsf{}& -1\%&
~~~--\textsf{}& ~~~--\textsf{}& -4\%\textsf{}& 4\%& 6\%\textsf{}&
14\%\textsf{}&
42\% \\
\hline
\multicolumn{11}{|c|}{December 6 2000 -- variant 2 \quad ($R_{{\rm h}{\rm o}{\rm t}} = 2.28 R_{\odot}$, $n^{{\rm 2}} V = 24.0\times 10 ^{{\rm 5}{\rm 9}}$ cm$^{-3}$)}\\
\hline
Cool component& 0.020& 0.160& 0.376& 0.710& 1.755& 1.343&
0.856& 0.439& 0.113&
0.034 \\
\hline
Hot component& 5.841& 3.094\textsf{}& 1.410\textsf{}& 0.770&
0.358\textsf{}& 0.066\textsf{}& 0.023\textsf{}& 0.007& ~~~--&
~~~-- \\
\hline
Nebula& 2.737& 0.989& 0.871& 0.754& 0.600& 0.202& 0.114& 0.074&
0.030&
~~~--\textsf{} \\
\hline
Total flux& 8.598& 4.243\textsf{}& 2.657\textsf{}& 2.234& 2.713&
1.611\textsf{}& 0.993\textsf{}& 0.520\textsf{}& 0.143\textsf{}&
0.034 \\
\hline
Observed flux& 8.662& 4.257& 2.682& ~~~--& ~~~--& 1.635& 0.944&
0.478& 0.120&
0.024 \\
\hline
Observational error& 0.064& 0.039& 0.023& ~~~--& ~~~--& 0.001&
0.001& 0.001& 0.001&
0.001 \\
\hline
Deviation $D  $ & -1\%\textsf{}& 0\%\textsf{}& -1\%\textsf{}&
~~~--\textsf{}& ~~~--\textsf{}& -1\%\textsf{}& 5\%\textsf{}&
9\%\textsf{}& 19\%&
42\% \\
\hline
\end{tabular}

\bigskip $^*$ Total flux = hot component + cool component + nebula, \\
$\hphantom{^*}$ deviation from observations $D = (\mbox{total
flux} - \mbox{observed flux} ) / \mbox{observed flux}. $
\end{sidewaystable}

\begin{table}
\caption{Ratio of the numbers of ionizing photons and of
recombinations at various times.}
\bigskip

\renewcommand{\baselinestretch}{1.52}
{\normalsize

\begin{tabular}
{|p{11cm}|p{1.5cm}|} \hline Date (state of the system)& $ \mu $
 \\
\hline 15.09.1999 (quiescence)&
1.12 \\
22.11.2000 (outburst, rise to the second maximum)&
$1.00 ^{+ 0.25} _{-0.21} $\\
06.12.2000 (outburst, close to the maximum) -- variant 1&
$0.94 ^{+ 0.24} _{-0.20} $\\
06.12.2000 (outburst, close to the maximum) -- variant 2&
$0.76 ^{+ 0.24} _{-0.20} $\\
\hline
\end{tabular}
}
\end{table}

\begin{table}
  \caption {Fluxes from various regions of the nebula (in units of 10$^{-12}$~erg$\cdot$cm$^{-2}$s$^{-1}$\AA$^{-1}$)}

 \bigskip

\begin{tabular}
{|l|c|c|c|}
\hline
Source &$U$& $B$&$V$\\
\hline \multicolumn{4}{|c|}{22.11.2000}\\
\hline
Entire nebula&1.983&0.717& 0.631 \\
Hot wind (second term)& 0.465 & 0.135 &
0.118 \\
Shock (third term)& 0& 0&
0\\
Cool region of the nebula&
 1.518 &
 0.582 &
 0.513 \\
   \hline \multicolumn{4}{|c|}{06.12.2000 -- variant 2}\\
\hline Entire nebula& 2.737& 0.989&
0.871\\
Hot wind (second term)&
 0.465&
 0.135&
 0.118\\
Shock (third term)& 0.657& 0.237&
0.209\\
  Cool region of the nebula&
1.615&
0.617&
0.544 \\
\hline
\end{tabular}
\end{table}
\end{center}

\clearpage
\begin{center}

FIGURE CAPTIONS

\end{center}

\vspace{0.5cm}

\medskip\noindent {\bf Fig.~1.}
The $UBV$ light curves of Z And during the 2000 outburst. Data of
Skopal et al. [\ref{Sk02}, \ref{Sk04}] (points) and our own
data (crosses) are shown. The first and second light
maxima are marked by dashed lines and the times for which the
ratios of ionizing photons and recombinations $\mu$ were calculated --
by arrows. The inserts show the behavior of the light near its maximum on a
larger scale.

\medskip\noindent {\bf Fig.~2.}
The light curve of Z And during the 2000 outburst, from Sokoloski et
al. [\ref{Sokol06}].

\medskip\noindent {\bf Fig.~3.}
Variation of the accretion rate for the solution with an increase
of the wind velocity from 20 to 30 km/s. The vertical dashed line
marks the time when the wind velocity changes.

\medskip\noindent {\bf Fig.~4.}
Contours of equal density and velocity vectors for six moments
during the outburst's development: (a) 12, (b) 42, (c) 70, (d)
103, (e) 125, and (f) 150 days after the start of the
computations. The hollow circle denotes the donor (the radius
corresponds to the donor radius).

\medskip\noindent {\bf Fig.~5.}
Contours of equal density and velocity vectors for the
two-dimensional computations. The hollow circle denotes the donor
(the radius corresponds to the donor radius). The distances are in
solar radii.

\medskip\noindent {\bf Fig.~6.}
Contours of equal density and velocity vectors in the region of
the shocks.

\medskip\noindent {\bf Fig.~7.}
The observed fluxes in the $UBVRIJHKLM$ wavebands.

\renewcommand{\figurename}{Fig.}

\clearpage
\begin{figure}
\includegraphics{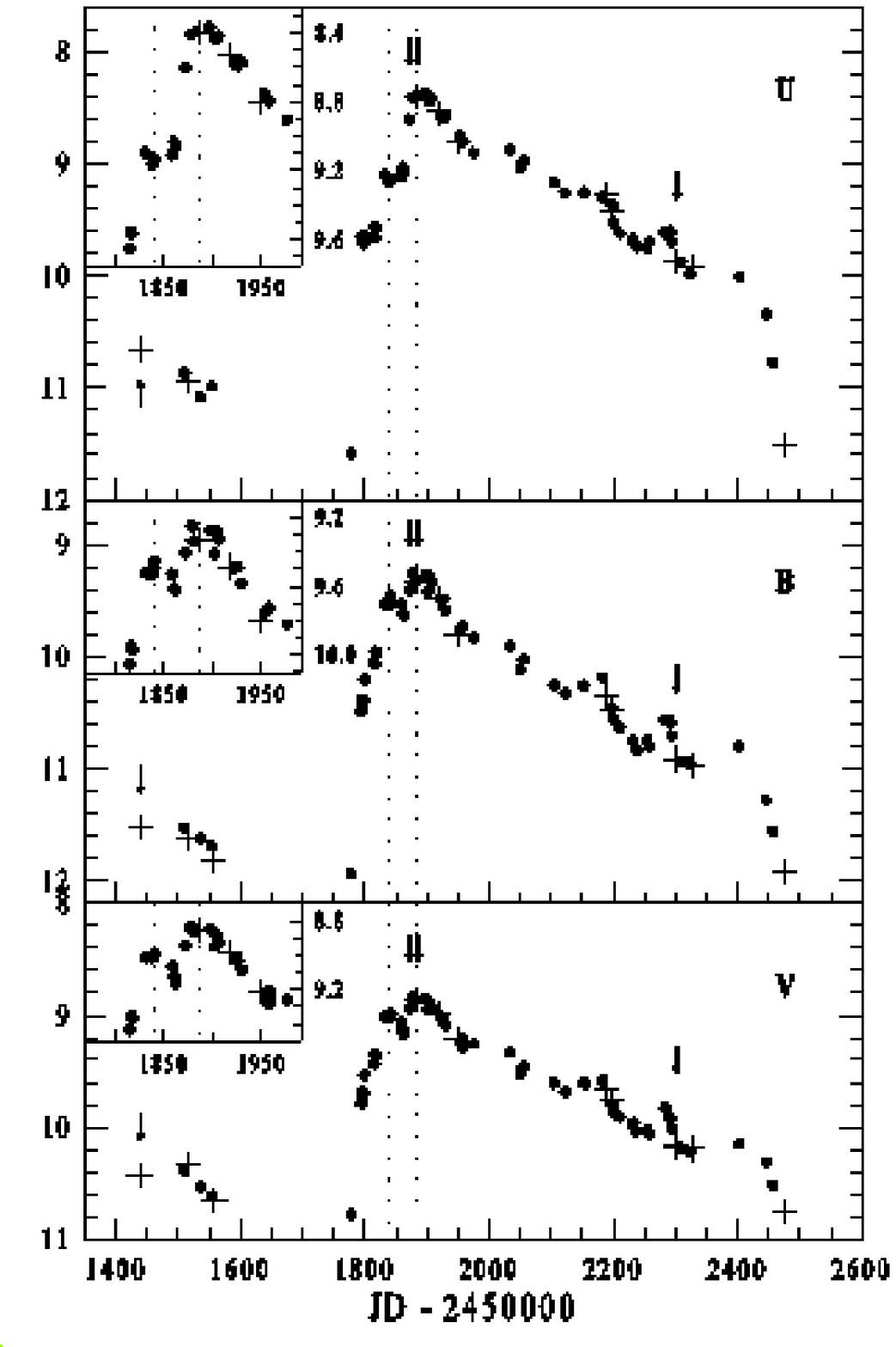}

\bigskip {\bf Fig. 1.}
\end{figure}

\clearpage
\begin{figure}
\includegraphics[scale=0.5]{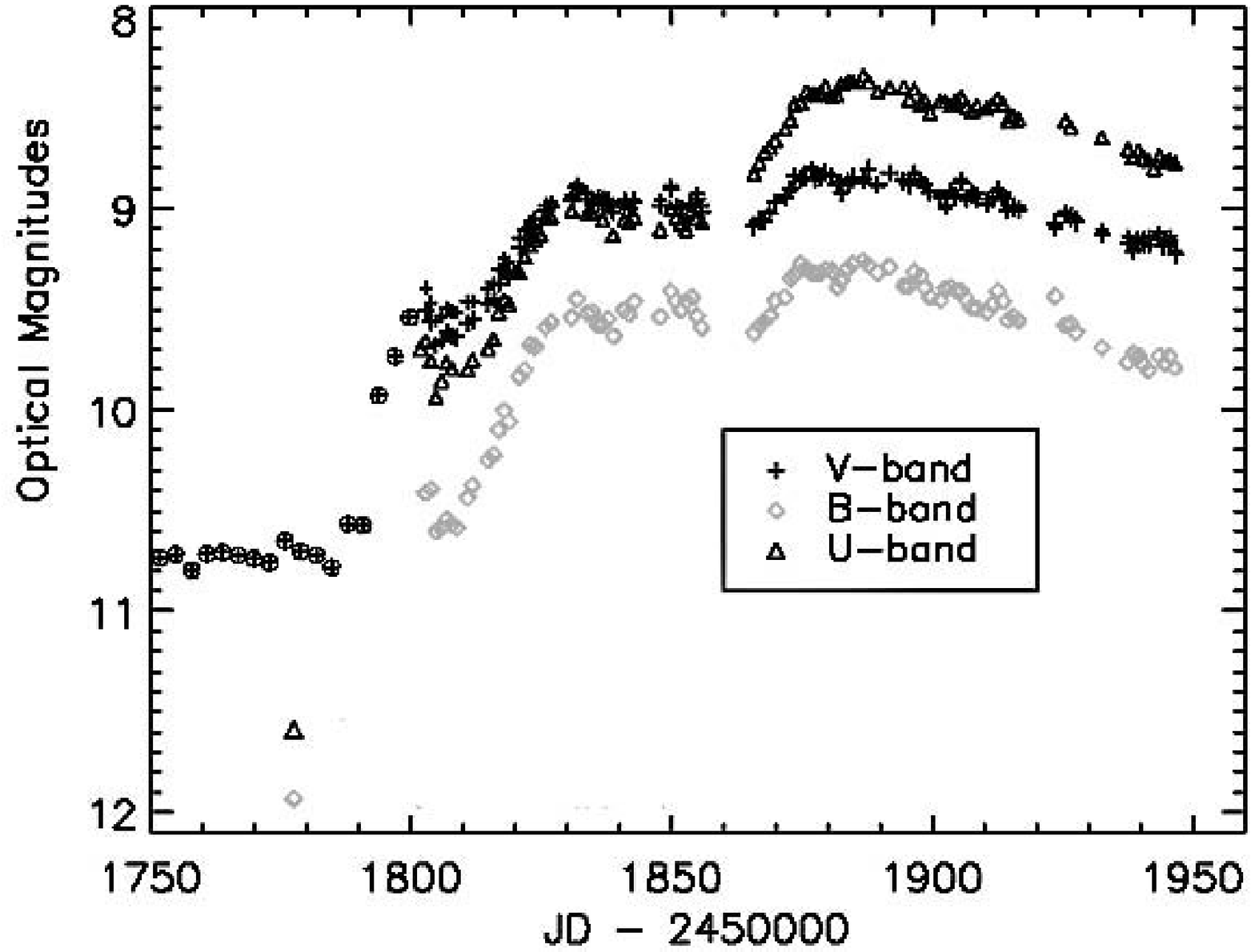}

\bigskip {\bf Fig. 2.}
\end{figure}

\clearpage
\begin{figure}
\includegraphics[scale=1]{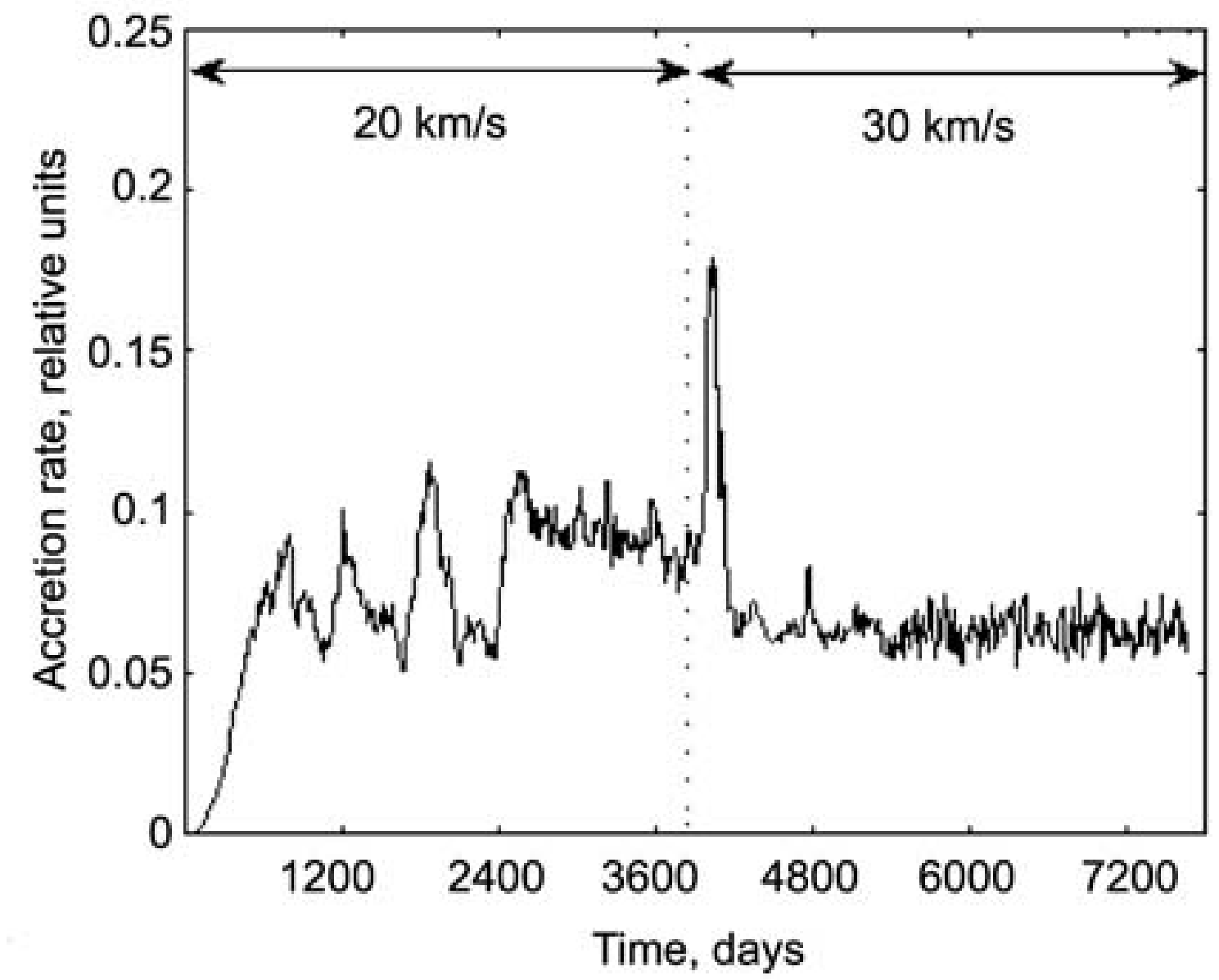}

\bigskip {\bf Fig. 3.}
\end{figure}

\clearpage
\begin{figure}
\includegraphics{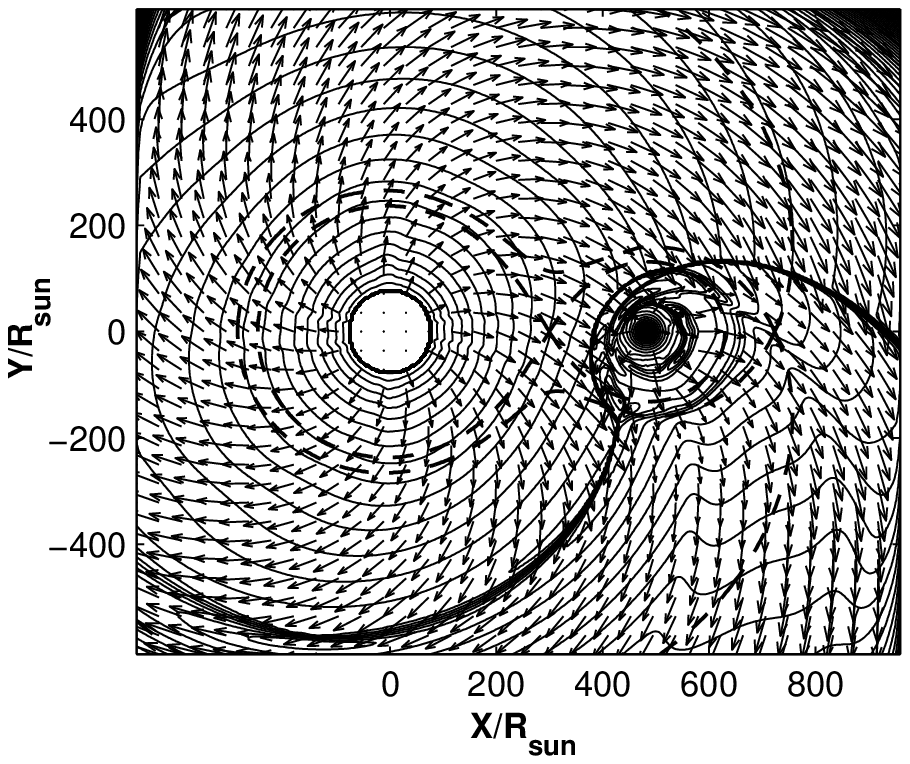}

\bigskip {\bf Fig. 4a.}
\end{figure}

\clearpage
\begin{figure}
\includegraphics{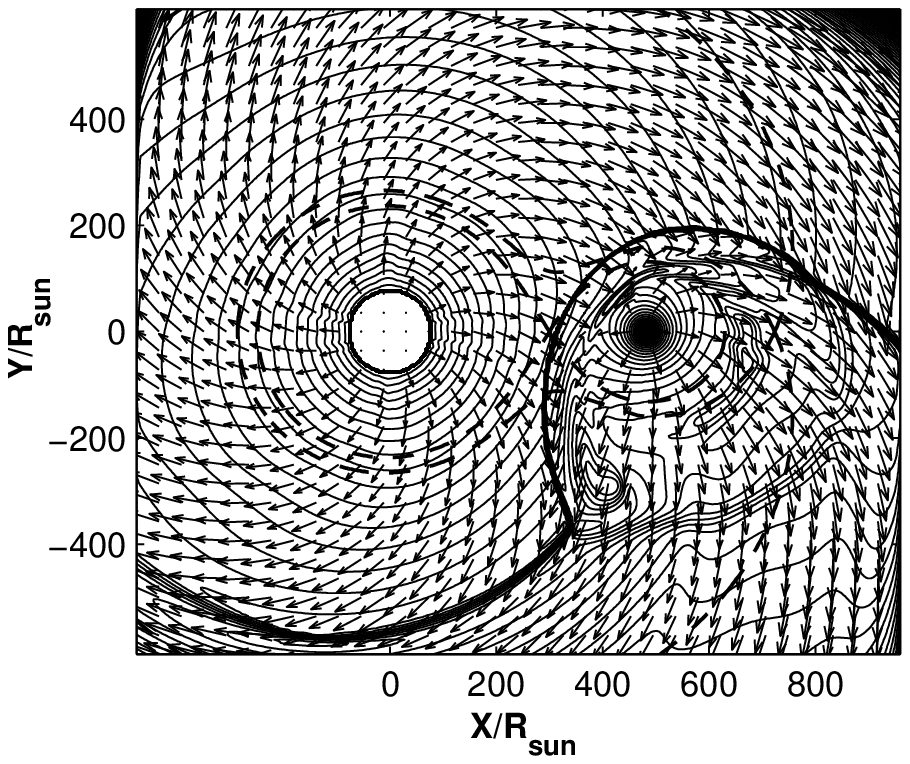}

\bigskip {\bf Fig. 4b.}
\end{figure}

\clearpage
\begin{figure}
\includegraphics{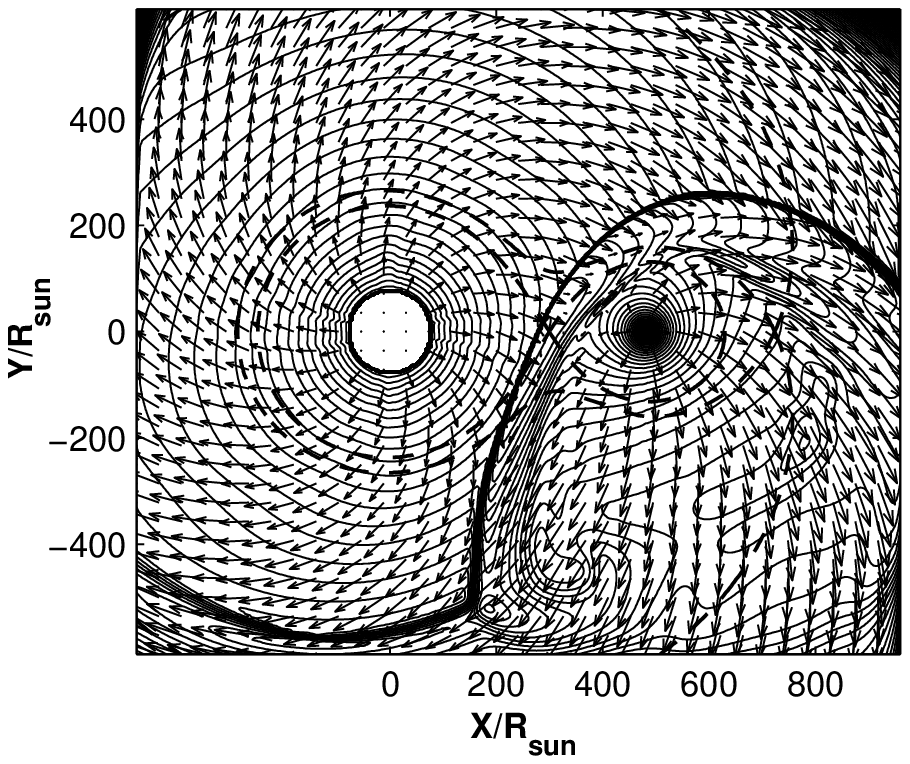}

\bigskip {\bf Fig. 4c.}
\end{figure}

\clearpage
\begin{figure}
\includegraphics{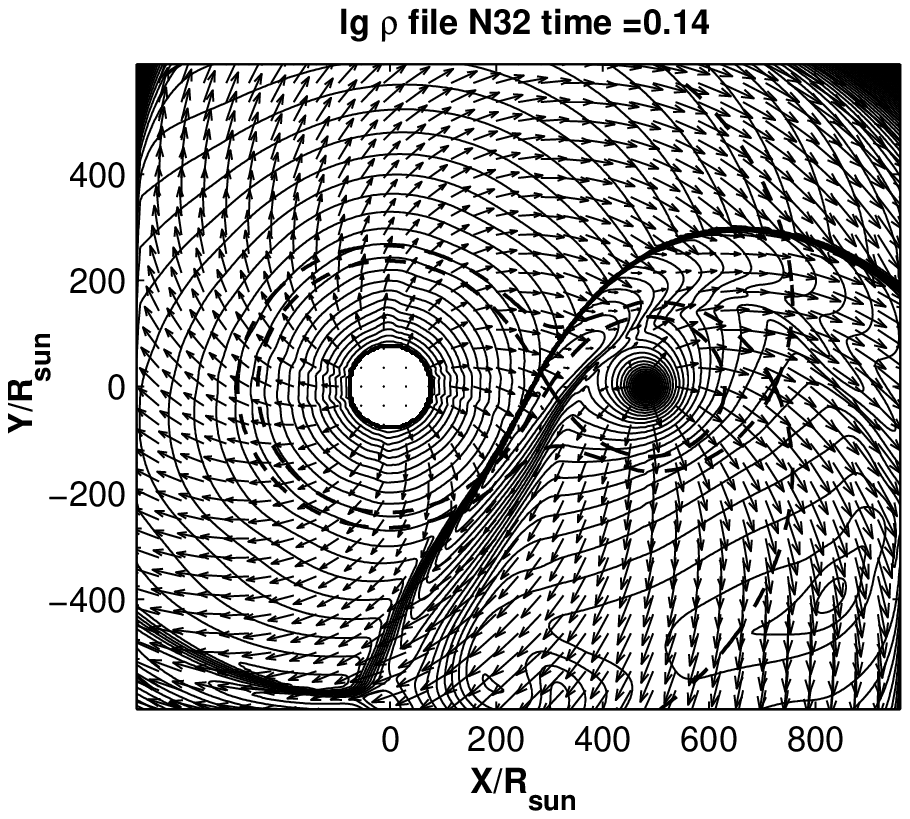}

\bigskip {\bf Fig. 4d.}
\end{figure}

\clearpage
\begin{figure}
\includegraphics{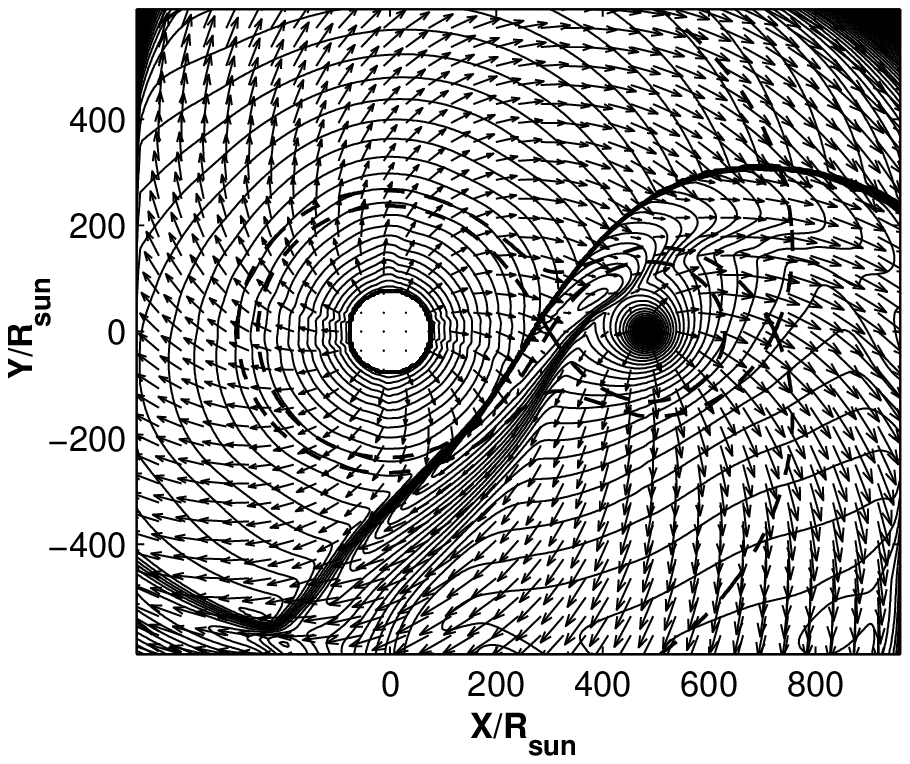}

\bigskip {\bf Fig. 4e.}
\end{figure}

\clearpage
\begin{figure}
\includegraphics{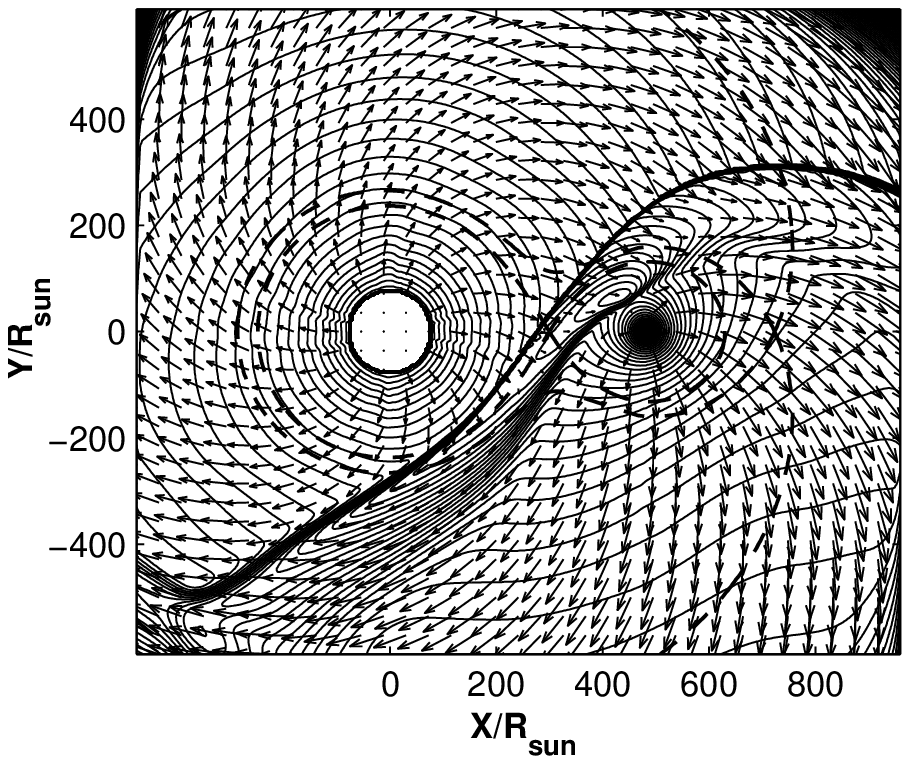}

\bigskip {\bf Fig. 4f.}
\end{figure}

\clearpage
\begin{figure}
\includegraphics{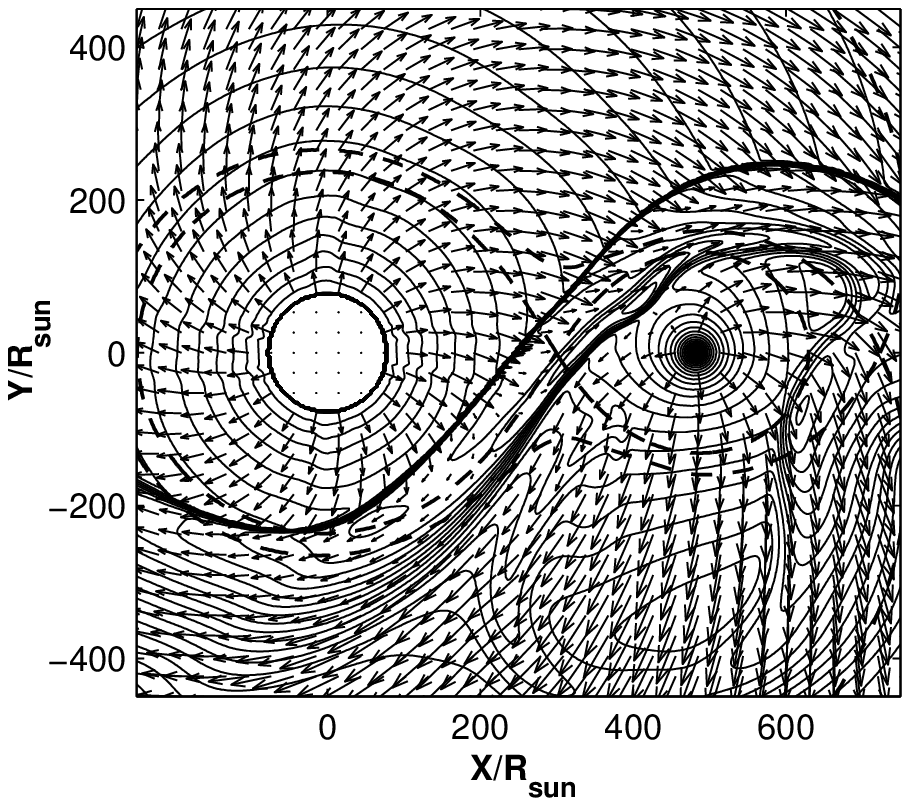}

\bigskip {\bf Fig. 5.}
\end{figure}

\clearpage
\begin{figure}
\includegraphics{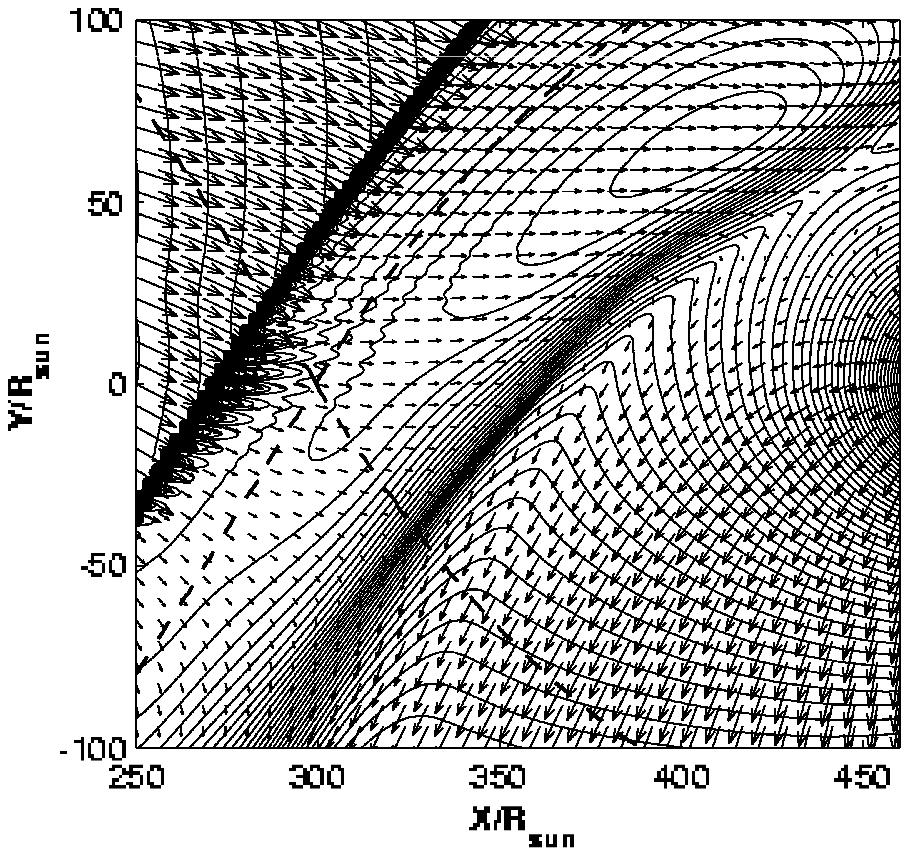}

\bigskip {\bf Fig. 6.}
\end{figure}

\clearpage
\begin{figure}
\includegraphics{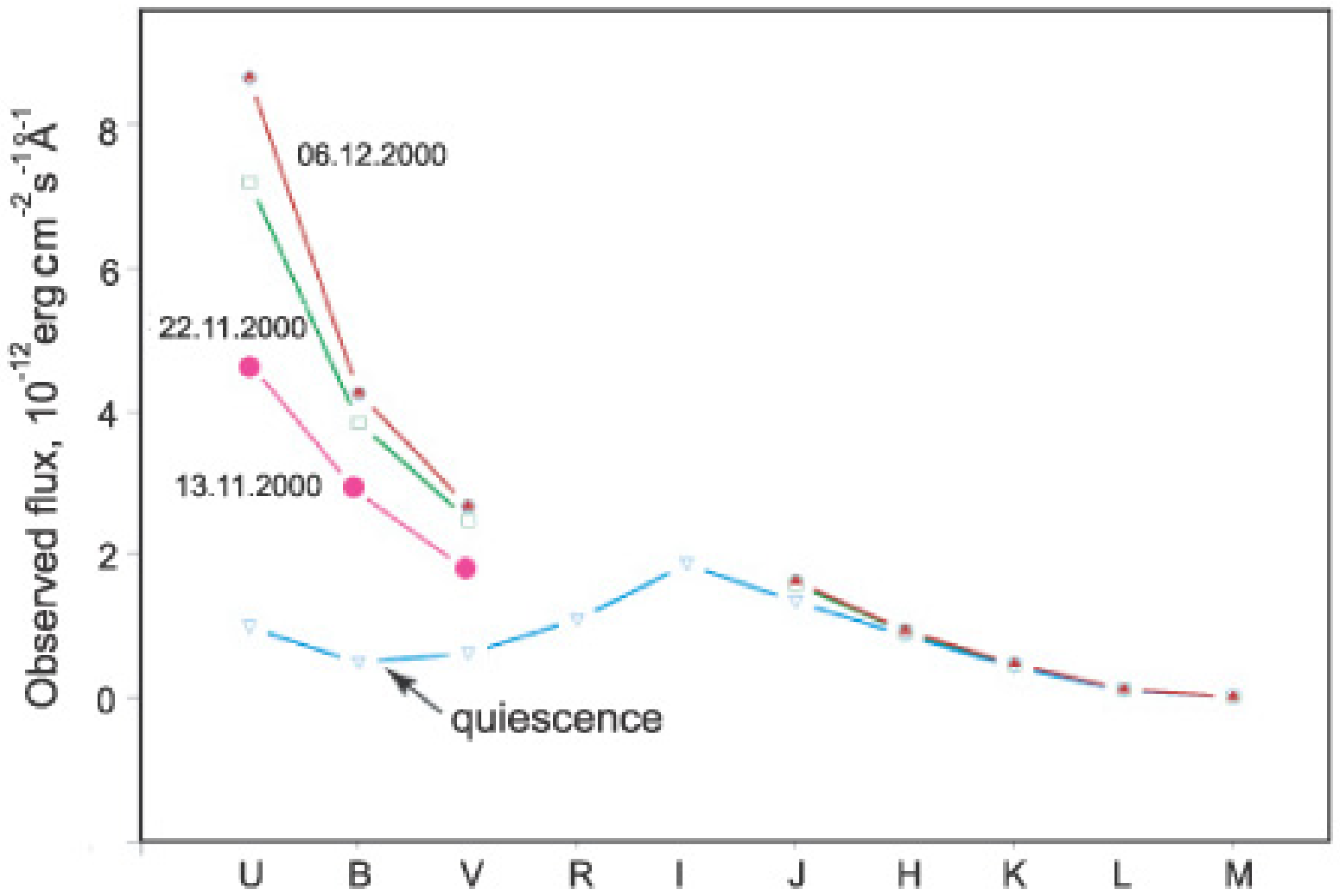}

\bigskip {\bf Fig. 7.}
\end{figure}

\end{document}